# Scalable Two-Minute Feedback: Digital, Lecture-Accompanying Survey as a Continuous Feedback Instrument

Armin Egetenmeier* and Sven Strickroth

Ludwig-Maximilians-Universität München, Munich, Germany
Email: armin.egetenmeier@ifi.lmu.de (A.E.); sven.strickroth@ifi.lmu.de (S.S.)
*Corresponding author

*Abstract*—Detailed feedback on courses and lecture content is essential for their improvement and also serves as a tool for reflection. However, feedback methods are often only used sporadically, especially in mass courses, because collecting and analyzing feedback in a timely manner is often a challenge for teachers. Moreover, the current situation of the students or the changing workload during the semester are usually not taken into account either. For a holistic investigation, the article used a digital survey format as formative feedback which attempts to measure student stress in a quantitative part and to address the participants' reflection in a qualitative part, as well as to collect general suggestions for improvement (based on the so-called One-Minute Paper) at two educational institutions. The feedback during the semester is evaluated qualitatively and discussed on a meta-level and special features (e.g. reflections on student work ethic or other courses) are addressed. The results show a low, but constant rate of feedback. Responses mostly cover topics of the lecture content or organizational aspects and were intensively used to report issues within the lecture. In addition, Artificial Intelligence (AI) support in the form of a large language model was tested and showed promising results in summarizing the open-ended responses for the teacher. Finally, the experiences from the lecturers are reflected upon and the results as well as possibilities for improvement are discussed.

## I. Introduction

Feedback is an essential aspect of successful teaching learning processes [1]. Most often the focus is on feedback for students, although teachers also need or want feedback. Universities use a variety of evaluation and quality assurance measures to support teachers and ensure high quality teaching. Institutionalized forms consist of standardized questionnaires (paper or digital) and rarely consider the students' context. In practice, these often only have a limited contribution to improving teaching due to the one-off nature of an end-of-term survey and its standardization [2, 3]. The aim of this kind of evaluation is usually the long-term assurance of teaching quality or the fulfillment of accreditation requirements. Hence, the gathered feedback is particularly relevant for subsequent cohorts [4]. For short-term adjustments in teaching, a more targeted, continuous survey over the term is required. Here, feedback such as students' current open questions or level of stress can also be helpful and incorporated into following teaching units. Particularly in the first semester, being able to react to student's needs and issues can ease the transition to university. It is therefore useful to supplement the end-of-term evaluation with additional, regular surveys on one's own course, so that it can be tailored to the current needs of the students and one gets quick feedback on (newly) used teaching methods [3]. In this way, teaching innovations can be consolidated, further developed, or discarded more quickly. However, courses in the first semesters are often attended by several hundred students at universities and making frontal lectures a last resort [5]. For this reason, social interactions, discussions and feedback become rather limited. Only a few people give direct feedback to the teachers about the course, as new students are more reluctant to give feedback and to suggest improvements [6]. Therefore, collecting feedback is complex and with a large number of responses, a (manual) evaluation can often only be done selectively or takes a significant amount of time [7]. Consequently, technological support is essential for scaling up and for use in digital or hybrid formats.

This article describes a digital, formative feedback approach that was used continuously during the term at two different educational institutions. It was implemented as a short, weekly survey and is referred to in the text as a two-minute feedback survey (2MF survey). The feedback is based on typical questions of the "Minute Paper" [8] with additional close-ended questions and aims to be completed within two minutes. The feedback received is intended to enable the lecturer to efficiently get feedback on the course and to gain insight into the learning progress and workload of the students. The following Research Questions (RQ) are investigated using three case studies: (RQ1) How do motivation to attend, workload and perceived stress affect the subjective understanding of being able to follow the lecture, how does this vary over the semester, and are there differences between educational institutions (or subject areas)? (RQ2) What free text answers do students give as feedback to open-ended questions? What content clusters can be formed and do they change over time? (RQ3) How well can large language models (i.e., ChatGPT) summarize the feedback given by students?

The contributions of this article are twofold: First, it is examined how this tool was used throughout the semester, what topics the feedback covers and how subjective factors of the students affect their assessment of their ability to follow the course. Only few studies have qualitatively analyzed the responses to explore common topics. This helps instructors to design and optimize surveys. The advantages of digital implementation for teachers are also discussed. Second, it presents results of a follow-up investigation on how large language models (LLM) can help to summarize feedback to meet the needs of large classes in a timely manner. This is particularly important to scale the feedback analysis and make it easier to grasp for teachers.

## II. RELATED WORK

Feedback methods that are easy to use and implement, such as hand signals or audience response systems, can give a first impression of the learning progress in the courses. They are, however, often not sufficient for a deeper insight and offer little opportunity to get students' suggestions for improvement. A suitable method for this is the "One-Minute Paper" (OMP) [8], which can support learners and teachers as formative feedback. Short questions to learners are used to make them reflect on the learning progress, to formulate open questions about the course and, if necessary, to collect general suggestions for improvement. The implementation includes a survey with two or three short reflection questions, usually at the end of a lesson, which allows for timely and concrete feedback [9]. Typical questions relate to content, materials, or specific points of interest to the teacher [9]. These may relate to the use of technology [10], teaching methods and style [7], newly learned concepts [11], or expectations of the course [6]. In literature, a combination of the questions "What was the most important thing you learned today?" and "What question(s) remained unanswered?" are often used [4, 10, 12]. The focus usually is on the learner's reflection on the learning content and thus on the learning progress of the group [4]. The teacher evaluates the written feedback and can address ambiguities, misunderstandings, or misconceptions [4, 13]. Responses about possible suggestions for improvement, can provide teachers with concrete ideas on how to better adapt the course to the group's needs [10, 11]. The uncomplicated feedback of the OMP has shown to encourage active engagement [4], even in large or mostly reserved groups [14], and to establish an atmosphere of trust [15]. Especially when used during a lecture, students can be re-engaged [9]. Choosing the right time and frequency to use OMP is still a point of dispute [16]. Previous studies have shown the greatest effect when used at the beginning of the term [12], or for targeted course feedback [6]. In contrast, this paper examines the effects, when the feedback is collected continuously throughout the whole semester.

Evaluating many written responses is time-consuming [8], especially when they are collected on paper [4]. Despite the scalability of a digital version for collecting the responses, there are only few studies dealing with purely digital implementations or investigating (continuous) use in mass courses [7, 10, 17]. Digital versions of the OMP seem to be mainly used in courses with a significant digital component or in fully virtual courses [11, 17, 18]. As technology becomes more important in teaching, face-to-face courses are also using digital OMP formats [18]. However, qualitative analyses of the responses to explore common topics is rare [11, 13]. This paper investigates the responses of a large course. Additionally, the (manual) analysis of a large number of responses to open-ended questions is still an open issue [19]. Here, automatic text summarization approaches [20] or upcoming (easier to use) LLMs such as ChatGPT might help and need to be investigated. The latter is addressed in this paper to provide practical insights for teachers and lecturers.

OMPs rarely consider students' current situation, such as the workload or perceived stress during the semester. For instance, a study found a correlation between self-rated workload and the received grades [17], but the information is not used within the semester (e.g. to support the students). Studies dealing with students' perceptions of stress [21, 22] often focus on the causal dimensions or on the use of external stress regulating help. Usually, extensive, established questionnaires from psychology are used as a basis, such as the Perceived Stress Questionnaire (PSQ) with 30 items [23]. There seems to be less focus on using the results to adjust a course. A digital implementation of the OMP combined with a short, quantitative part for a rough assessment of the students' current situation during the semester has not yet been investigated in the literature.

## III. STUDY DESIGN AND METHODOLOGY

An experimental case study was conducted with the aim to implement and evaluate a scalable, digital feedback process. The 2MF survey allows continuous, low-threshold collection and quick analysis for reflection, adaptation of teaching and provides a first insight into the student workload, even in large courses. At the same time, the feedback will allow students to reflect on the course content. The data were collected in parallel at two different German educational institutions during the winter term 2022/2023 by administering the 2MF survey weekly in an introductory computer science course at Ludwig-Maximilians-Universität München (LMU Munich, referred to as Uni in the following) and an introductory mathematics course at Aalen University of Applied Sciences (UAS). To investigate the feedback topics, an inductive categorization analysis [11] was carried out by one researcher on the given responses. Starting with general topics based on random selected answers (e.g. "teacher related" or "organization"), the researcher used an iterative process to add emerging responses in categories such as "lecture content", "self-reflection" and "generic answers", or details due to repeated mention (e.g. "slides/script" or "exercises"). Thus, the answers could fall into several categories. The topics were used in both case studies. The closed questions were analyzed for statistical correlations. The feedback of the students and the identified topics are then used as a basis for the investigation of the quality regarding completeness of the summarized feedback of three weeks using ChatGPT.

The 2MF survey is structured in two parts: The first part consists of six closed questions about the student's situation ("I feel stressed." and "I feel overwhelmed by my studies."), motivation ("I am motivated to attend the lecture."), and understanding ("I can follow the content of the lecture well.") each on a five-point Likert scale with –1 for abstention, 1 for disagreement, 5 for agreement; and two yes/no questions regarding attendance ("I attended/watched the lecture/ …exercises last week."). The second part is strongly based on the OMP and includes a question about unclear content, a question about suggestions for improvement, and a reflection question about what students liked most about the lecture last week. The survey was implemented in the e-assessment system Generic Assessment & Testing Environment (GATE) [24] (Uni) and as a protected website (UAS) with Single Sign-on so that the data could be pseudonymized and collected once a week per person. The front end for the students is kept simple and only shows a form with the survey questions in order to have the lowest possible response burden. In contrast to surveys for reflection and evaluation, the first two questions (stress and feeling overwhelmed) asked about the overall subjective assessment of the student situation. The teacher receives the feedback in a dashboard, which contains all collected free text answers and pie charts for each closed question per week (see Fig. 1), supplemented by an overview of the quantitative results over the last weeks in line charts (see Fig. 2).

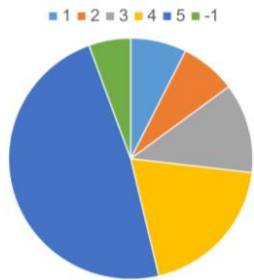

Fig. 1. Detailed evaluation for "understanding" Uni, week 44 (1=do not agree; 5=fully agree; −1=abstain).

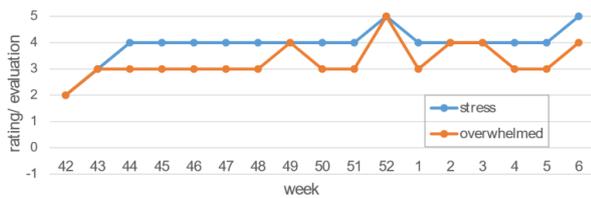

Fig. 2. Semester overview (median) for "stress" and feeling of being "overwhelmed".

## IV. CONTEXT OF THE CASE STUDIES

The first case study was conducted in the compulsory first semester course "Fundamentals of Business Mathematics" at Aalen University of Applied Sciences. The course is divided into a lecture part and two separate exercise groups, each of which is led by an experienced lecturer. There was no supplementary tutorial offer by student teaching assistants. Around 80 students were enrolled in the course. The lectures took place every second week with irregular durations, whereas the exercise sessions were offered almost every week of the semester. The individual lessons were either face-to-face or virtually (via video conferencing tool). The voluntary 2MF survey was explained in the first lecture, mentioned several times in the following lectures and regularly mentioned in (digital) announcements. The information and link were provided in the Learning Management System (LMS). Questions raised in the responses were taken up, repeated and answered at the beginning of the next lecture. Suggestions were commented on and implemented as far as possible.

The second case study took place as part of the first semester course "Introduction to Programming" at LMU University. This course introduces fundamental concepts of computer science using the programming language Java and consists of four hours of lectures (usually 2 hours theory and 2 hours practice session) and two hours exercise sessions each week. The lecture was attended by around 900 students and offered in hybrid form. Recordings were made available online. A reminder for the weekly survey was displayed (using a QR code link) at the end of the last lecture of each week. In addition, there was a link in GATE that was displayed above the assignments if the survey had not yet been carried out in the current week. Raised questions and comments were taken up and answered as far as possible in terms of time within the lecture.

## V. FINDINGS OF THE CASE STUDIES

### A. Descriptive Evaluation

In total 274 (Uni) and 14 (UAS) distinct students provided feedback in the two courses. A total of 726 (Uni) and 18 (UAS) responses were recorded from these individuals (2.6 and 1.3 responses per person, respectively). A total of 486 free text answers were received at the Uni and 24 at the UAS. Whereas suggestions for improvement slightly pre-dominate at the Uni (approx. 30 more answers), the answers at the UAS are distributed more or less evenly (between 7–9 answers; see Fig. 3, bottom). There are 3 people at UAS who participated in the survey repeatedly (max. twice). A total of 140 people from Uni repeatedly took part in the survey (max. 14×, 25 people more than 7×). It should be noted that the feedback is distributed over almost all days of the term (Uni) and was also received during the lecture session. The feedback from the UAS all came afterwards. At the Uni, consistent participation in the 2MF survey can be seen throughout all weeks (see Fig. 3, top), whereas at the UAS there are individual weeks of no participation despite taught courses. Over time, a small core (of 25 students) developed at the Uni who gave feedback almost continuously. After calendar week 50 (before the last week of the course), no more feedback was received at the UAS. According to the 2MF survey, a total of 673 (Uni) or 16 (UAS) students confirmed having attended the lecture in the corresponding last week and 445 (Uni) or 17 (UAS) for the exercises during the term. There are 31 responses (1× UAS, 30× Uni) from students who neither attended the exercise session nor the lecture in the corresponding last week.

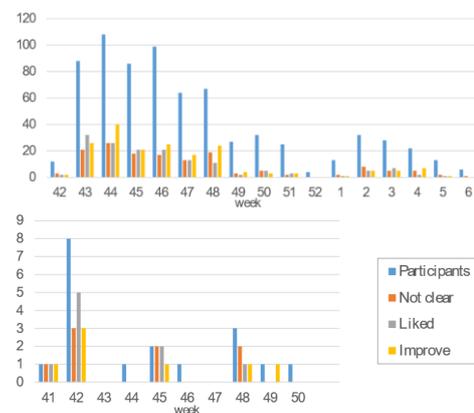

Fig. 3. Number of responses per feedback category and calendar week for Uni (top) and for UAS (bottom).

Table 1 shows an excerpt of the data relating to the first nine weeks for Uni (starting at calendar week 42) and ten weeks for UAS (starting at calendar week 41). In addition to the number of participants, the number of responses on the questions "What remained unclear", "I liked" and suggestions for improvement, and four identified categories of the categorization analysis are also displayed: The most common content of the coding of the UAS falls into the areas of "lecture content", "structure/procedure" (organization) and "generic answers" (general), as well as for Uni in "lecture content", "exercises" and organization resp. general. In both cases, there is a large coverage of the categories. Allocations of the answers to multiple categories is possible.

Table 1. Number of responses and results of the free text coding for Uni (first number, left) and UAS (second number, right) of the first 9 or 10 weeks in the respective semester (calendar week 41 to 50)

| # | 41 | 42 | 43 | 44 | 45 | 46 | 47 | 48 | 49 | 50 |
|---|---|---|---|---|---|---|---|---|---|---|
| Participants | –/1 | 12/8 | 88/0 | 108/1 | 86/2 | 99/1 | 64/0 | 67/3 | 27/1 | 32/1 |
| Unclear | –/1 | 3/3 | 21/0 | 26/0 | 18/2 | 17/0 | 13/0 | 19/2 | 3/0 | 5/0 |
| Liked | –/1 | 2/5 | 32/0 | 26/0 | 21/2 | 21/0 | 13/0 | 11/1 | 2/0 | 5/0 |
| Improve | –/1 | 2/3 | 26/0 | 40/0 | 21/1 | 25/0 | 17/0 | 24/1 | 4/1 | 3/0 |
| Lecture Content | –/0 | 1/4 | 13/0 | 23/0 | 17/2 | 13/0 | 7/0 | 19/2 | 3/0 | 6/0 |
| Organization | –/0 | 1/3 | 12/0 | 37/0 | 7/3 | 13/0 | 8/0 | 6/1 | 3/0 | 3/0 |
| Exercises | –/0 | 0/0 | 2/0 | 24/0 | 5/2 | 20/0 | 18/0 | 11/0 | 5/0 | 2/0 |
| General | –/0 | 2/3 | 16/0 | 19/0 | 12/0 | 11/0 | 3/0 | 9/1 | 1/0 | 3/0 |

### B. Analysis of Time and Content

The first part of the 2MF survey is about the students' assessment of their current state of health. This part was answered more frequently during the term than the open-ended questions. The number of participants steadily decreased over the term (see Fig. 3 and Table 1). While no linear correlation can be found for the UAS, there is a statistically significant linear correlation with time for the Uni (Pearson's $r(14) = –0.88$, $p < 0.001$). This is also reflected in the submission of weekly assignments ($r(8) = –0.97$, $p < 0.001$).

Initially, placeholder symbols such as "-" were used frequently as responses (only Uni). There are also more generic comments ("general") in the first few weeks, such as "everything", "nothing", "yes", "no", which were used much less frequently towards the end of the term. Questions about lecture content ("unclear") were asked throughout the whole term (see Table 1). Looking at the first six weeks of the semester (Uni), the median comment length is 6 words and increases towards the end of the term (last six weeks, median 9.5 words). At the UAS, the median comment length for the entire term is 8.5 words. Criticism is usually formulated in more detail. There is a particularly large amount of feedback on the organization of the course at the beginning of the term at the Uni (37× in week 44, see Table 1). The vast majority referred to technical and organizational issues that arose during the practice sessions. In this case, the lecturer (Uni) was only able to give a regular lecture starting from week 45 due to illness, so in week 44 there was only a practice session with a substitute teacher without a theory lecture. The number of responses also shows that there were no more lectures at the UAS from week 51 onwards (but only exercises).

### C. Special Features and Content (Meta) Analysis

A closer look at the content reveals that there are repeated entries, i.e. a statement was copied into all free text fields to emphasize it (Uni, week 47 and 48). These were of a more critical nature (e.g. "Do you actually read the feedback?"), combined with a suggestion for improvement. The feedback (Uni) indicated problems at an early stage, e.g. a tense mood in a lecture (week 44) or infrastructural challenges (missing power sockets, week 45) as well as technical and didactic subtleties ("Repeat questions of the local audience for online participants", week 50). Problems with accompanying exercises could be identified and tackled this way (week 45). Some answers were also conspicuous, which extended the actual question in an interpretative manner ("[This course] is great, I'm stressed out by Analysis and Algebra :(" Uni, week 45, in category "unclear", or "So far I only had positive emotions about the lecture" Uni, week 46, in category "improve"), or completely different points were taken up ("everything was understandable" Uni, week 48, "unclear"). Some comparisons were made with other courses and these were related to the course. Interestingly, the responses included (self-)reflections by individual students on their own work ethic ("My mistake because I didn't attend the lecture regularly" Uni, week 3, "unclear"), or general insights regarding live programming submissions ("Forty submissions from almost 900 participants [...] are not a good result [...] a submission rate of 20 % [is] pathetic..." Uni, week 48, "improve").

### D. Data Linking and Correlation (Uni Only)

When the responses to the closed questions are analyzed for correlations, a large statistically significant linear correlation between "stressed" and "overwhelmed" could be found ($r(720) = 0.75$, $p < 0.001$). There is also a slightly negative linear correlation between "stressed"/"overwhelmed" and "could follow" ($r(704) = –0.20$, $p < 0.001$ and $r(706) = –0.27$, $p < 0.001$). Furthermore, "overwhelmed" seems to have a slightly negative effect on "motivation" ($r(704) = –0.14$, $p < 0.001$), whereas "could follow" has a positive effect ($r(699) = 0.46$, $p < 0.001$).

A total of 713 students (241 of whom provided feedback) consented to the use of their data for research. A statistically significant difference in the number of exercise sheets submitted in the LMS could be found: The median number of submissions from students who provided feedback is 7, otherwise 4 (UTest: $U = 39.938$, $p < 0.001$).

## VI. REFLECTION ON THE RESULTS BY THE TEACHERS

In the UAS case study, the number of responses seems to indicate that this form of continuous feedback is not (yet) desired on a broad basis or that its added value is not (yet) recognized. Nevertheless, from the teacher's point of view, the feedback on content that was not understood by students provided a good starting point for repetition in the next session. Issues did not have to be laboriously asked or anticipated from the exercises. The added (anonymous and) direct communication channel is welcome despite the low level of participation, as the existing alternatives (e.g. forum posts, email inquiries) were used even less. Due to the lack of data, insights into the general workload of the individual students could not be evaluated.

In the Uni case study, there were several responses that offered exciting insights from the teacher's point of view, e.g. repeating questions from the local audience for the live stream. This shows that students apparently do not dare to address this directly in the chat. Also, not all questions or comments could always be addressed, e.g. regarding the speed of the lecture or material discussed several weeks ago. It remains unclear why questions on the content were not discussed in the exercise sessions—maybe no satisfactory answer was given there. Nevertheless, such feedback is not only time-consuming for students, but also for teachers. Dealing with it costs extra time, which is worth it, but it would be nice to reduce the time needed even further.

## VII. ARTIFICIAL INTELLIGENCE (AI) SUPPORT

The third case study involves the use of AI to get the key insights from a large amount of responses to improve scalability. After the term and the manual coding, the responses of weeks with the most answers (Uni, week 43 and 44) and a high number of words per answer (Uni, week 47) were summarized using the LLM ChatGPT (May 24 Version on June 1, 2023). Each free text field of the 2MF survey (unclear, liked, improve) was investigated individually and the summarization was conducted three times for not relying on a single result. As all responses are German, German was also used for the AI prompts. The goal was to investigate whether this can be used to quickly summarize (many responses) and find a proper categorization of main topics (see Section III).

Table 2. Comparison of the number of the original text responses and words (left) with the results of ChatGPT (right)

| Week | item | # responses orig./LLM | % | # words orig./LLM | % | # topics LLM |
|---|---|---|---|---|---|---|
| Week 43 | Unclear | 21/6–8 | 38% | 146/70–80 | 55% | 6 |
| | Liked | 32/6–13 | 41% | 270/60–90 | 33% | 8 |
| | Improve | 36/8–10 | 28% | 196/60–70 | 36% | 7 |
| Week 44 | Unclear | 26/8–12 | 46 % | 561/60–100 | 18% | 4 |
| | Liked | 26/7–10 | 38% | 167/60–80 | 48% | 4 |
| | Improve | 40/7–10 | 25% | 1133/50–100 | 9% | 3 |
| Week 47 | Unclear | 13/10–12 | 92% | 337/100–120 | 36% | 11 |
| | Liked | 13/10–11 | 85% | 214/70–100 | 47% | 10 |
| | Improve | 17/16–17 | 100% | 783/140–170 | 22% | 16 |

Note: Approx. values due to the generation of three answers from LLM. Percentage is based on the highest number of LLM result.

First, the summarization of the responses was investigated using the prompt "Summarize the main points of the following answers in bullet points." (translated). The median number of answers decreased from 26 to 11, the average number of words from 270 to approx. 100 (see Table 2). So, in most cases the number of words could be reduced by more than half without losing the essence of the answers: Responses like request for more testing options, emphasized as a copied response (see Section V-C) were still in the summary as well as the "desire to speak more slowly" (single response) or "difficulties with the practice session" (multiple responses). Thus, ChatGPT summarized all the relevant information. No new topics were introduced due to hallucinations.

Next, the focus was on categorization of the results using the prompt "Summarize the most important points of the following statements in bullet points in categories, use as many suitable categories as necessary. Also provide a concise, bullet-type overview of the most important, possible main categories" (translated). Besides a comparison with the first summary, this prompt should offer an even more reduced form of the main topics, which can be compared to the developed categorization (see Section III). As expected, the summary produced similar results with minor variations (such as "Please speak more slowly") as the response mentioned above—easy to understand and useful for teachers. However, the categorization of the LLM showed interesting results: Some categories were similar to the general categorization (e.g. "practical orientation" or "communication and interaction"), while others were mostly ignored by the LLM (such as generic answers or placeholder symbols). In most cases, the LLM offered more details in the topics. For instance, the LLM used "Sympathy and positive characteristics of the lecturer" or "pace and motivation of the lecturer" instead of a general "lecturer related".

Finally, the prompts above were used on responses on the course of the UAS (cf. Table 1, week 42). The number of comments and words was identical or increased with the use of the AI. The already low number of responses encouraged the LLM to elaborate further. The categorization showed similar results as above.

## VIII. DISCUSSION

The development of clear and quickly answerable questions for the 2MF survey was difficult, especially with regard to subjective assessments. The selection of questions quickly raised ethical implications. To increase reliability, the use of established and tested questionnaires is desirable, but are often quite long and come from psychology [23]. Hence, one is dealing with health data that are under extra protection of the General Data Protection Regulation. Also, dealing with such information is tricky: If, for example, data of a student is conspicuous, should be intervened to provide help or is addressing the issue also inappropriate, since lecturers cannot make a diagnosis. Hence, a few self-developed questions were used here. Asking for suggestions for improvement is also not unproblematic as it raises students' expectations that a change will then occur [9]. Furthermore, the case studies show that there are apparently only few incentives for (continuous) participation. The added value of participation (e.g. problem solving) is not recognized by all students, although questions and comments were answered timely (see Section V-D).

The implementation as a digital 2MF survey provides the desired, simplified collection of feedback for the lecturer and creates a more flexible feedback option for the students. Not only could more students be reached, such as non-participants from the last week, but ubiquitous feedback was also given (see Section V-A, [16]). Nevertheless, in relation to the respective course sizes (UAS: 80, Uni: 900 people), the number of responses is rather low overall. In relation to the group sizes considered, the 2MF survey reached around 17.5 % (UAS) and 30.4 % (Uni), but not consistently. Such low numbers have already been identified in interactive tools e.g. for participation in discussion forums [25] and seem to be a general problem (at least in Germany). On first sight, it may seem questionable to display the QR code at the end of the lecture, as entering free text answers is not ideal for smartphones. However, today's students are quite experienced in typing on smartphones, especially for short, chat-like posts, due to the intensive use of messengers etc. Also, the feedback was given at different points in time and not only during the lectures. The feedback (Uni, Fig. 3) in weeks 51, 52, and 1 stands out, because there are no events at the universities at the turn of the year. Nevertheless, more than 20 people gave feedback. It is interesting why there is a peak for stress and feeling of being overwhelmed (see Fig. 2).

The study shows that the initial number of responses decreases as the term progresses and levels off at a constant but low level [12]. There were an increased number of responses at the beginning of the term and especially in the case of (acute) problems relating to the course (see Table 1, Uni, week 44). Other communication channels (mail, forums, etc.) were not used to the same extent, which supports the use of a low-threshold feedback offer [12]. In particular, more engaged students seem to give (and demand) feedback, which is indicated by the higher number of submissions. The positive correlation of feedback and submissions on learning outcome is also seen in other research [14, 17]. A significant change in stress over the term (RQ1) can hardly be proven because there is too little feedback (UAS) or the evaluation of individual data (e.g. perceived stress level) leads to ethical implications. Nevertheless, the data provides a first impression of the context (at least for the Uni). The linear correlations between feeling of being overwhelmed, stress and the perception of "could follow" also suggest this (see Section V-D). An influence of the workload (Fig. 1) on the delivery of solutions is not recognizable. The results and clusters of the categorization analysis are consistent with other studies [11]. The categorization analysis (RQ2) nevertheless revealed surprising aspects such as the self-reflection of individual students or the comparison with other courses. This seems to strengthen the authors' assumption that evaluations of courses depend to some extent also on other courses students attended and that a holistic view of the student group therefore makes sense.

Using an LLM (here ChatGPT, RQ3) has already shown to be a good way of summarizing and providing a very good overview of the feedback, especially with a larger number of responses. The amount of feedback from the case studies was still manageable "by hand" (even in the Uni case study). If more feedback is provided by students, the likelihood of missing important responses due to teacher skimming is likely to increase. The main aspects of the free text responses were found by the LLM with a simple prompt instruction (Section VII)—even single (important) responses were honored in the summary. The LLM seems to collect requests and informs about all critical findings like difficulties or issues (Uni, week 44), which seems to be suitable for lecturers. In addition, detailed categories could be discovered by the AI without any specification (e.g. "motivation of the lecturer"). Yet, some aspects (e.g. self-reflection) remained hidden. Using AI on a small number of comments seems to offer limited benefit. These results suggest that the use of LLM can be a starting point for identifying relevant topics (from scratch). However, it should be used and evaluated with caution, as some categories may remain "unseen" by the AI. In this case a scientific evaluation with the categorization analysis comes in handy. From a teacher's point of view, this easy-to-use AI black box already offers a practical application for summarizing during the term—without hallucinating on topics.

Surveys of this type are limited by the self-selection of the participants and their willingness to provide feedback. For non-participants, no conclusions can be drawn about the reasons for the lack of feedback, nor can their feedback responses be predicted. Therefore, no feedback does not mean that there are no issues. Another limitation can be the content analysis itself. The inductive categorization was only carried out by one of the authors at this state, which is why a subjective view cannot be ruled out. Also, due to the rather low participation and the ethical aspects, an individual assessment of the students' workload is not possible.

## IX. Conclusion

In this article, a low-threshold, digital offer for continuous feedback called two-minute survey during the semester at two educational institutions was proposed and analyzed. This survey consists of open-ended and closed questions asking for further information on the context of the students such as workload and motivation. The digital implementation enables a scalable, ubiquitous and efficient way of data collection and evaluation, which enables formative feedback even in mass courses with little effort. There is added value for students and teachers alike, with manageable effort. The study results show that formative feedback for teachers can be used to better adapt teaching to the student group, especially at the beginning of the semester. Feedback on unclear teaching content not only allows learners to reflect, but also provides a suitable basis for in-class discussion and a starting point to revising the content or teaching format. In particular, the digital collection and evaluation enriches these options enormously, especially regarding learning analytics (visualizations in dashboards). The exemplary use of an LLM showed to be suitable to summarize the open-ended feedback for teachers, even if the used AI is a black box. It facilitates scaling even with a large number of responses despite issues when there are only few responses to summarize. This low-effort use of AI to evaluate (continuous) feedback can address concerns of teachers about being overwhelmed by feedback in large classes. Hence, the proposed approach shows how teachers can use such feedback approaches also in their large classes efficiently.

There is a need for further research on several levels: This primarily concerns increasing the regular participation in the surveys, whether through external incentives (e.g., gamification) or dashboards for students for self-reflection (e.g., workload and group comparison). Next, even if ChatGPT already showed a promising performance, more research on automatic summarization or clustering of the open-ended answers is desirable, so that central points can be seen quickly regardless of the number of responses. Furthermore, knowledge of the content of the feedback can be a basis for developing technological enhancements (e.g. chat bots) that can provide immediate responses to students (and teachers). Also, there could be proactive notifications to the lecturer based on the responses if specific aspects are mentioned. In this case, the categorization can be used as a starting point for implementing personalized alerts in advance.

CONFLICT OF INTEREST

The authors declare no conflict of interest.

AUTHOR CONTRIBUTIONS

Armin Egetenmeier conducted the content analysis and executed the case study involving AI support; Sven Strickroth came up with the ideas for the analysis and implemented the prototypes. Both authors contributed equally to the analysis of the case studies and the writing on the paper. All authors had approved the final version.

FUNDING

This research is part of the project AIM@LMU funded by the German Federal Ministry of Education and Research (BMBF) under the grant number 16DHBKI013. The responsibility for the content of this publication lies with the authors.